\documentclass[structabstract]{aa}  
\usepackage{graphicx}
\usepackage{txfonts}
\usepackage{natbib}
\def\LSXI{\mbox{LS~III~+46~11}}
\def\LSXII{\mbox{LS~III~+46~12}}
\def\BXC{\mbox{Berkeley~90}}
\def\teff{\mbox{$T_{\rm eff}$}}
\def\ebv{\mbox{$E(4405-5495)$}}
\def\rv{\mbox{$R_{5495}$}}
\def\logd{\mbox{$\log d$}}

\def\AV{\mbox{$A_V$}}
\def\mum1{\mbox{$\mu$m$^{-1}$}}
\def\chir{\mbox{$\chi^2_{\rm red}$}}
\newcommand{\HeI}[1]{\mbox{He\,{\sc i}~$\lambda${#1}}}
\newcommand{\HeII}[1]{\mbox{He\,{\sc ii}~$\lambda${#1}}}
\newcommand{\CIII}[1]{\mbox{C\,{\sc iii}~$\lambda${#1}}}
\newcommand{\NIII}[1]{\mbox{N\,{\sc iii}~$\lambda${#1}}}
\newcommand{\NIV}[1]{\mbox{N\,{\sc iv}~$\lambda${#1}}}
\newcommand{\OIII}[1]{\mbox{O\,{\sc iii}~$\lambda${#1}}}

\begin{document}
   \title{The little-studied cluster Berkeley 90}
   \subtitle{I. LS III +46 11: a very massive O3.5 If* + O3.5 If* binary}

   \author{J. Ma{\'\i}z Apell{\'a}niz\inst{1}
           \and
        	  I. Negueruela\inst{2}
        	  \and
	          R. H. Barb\'a\inst{3}
	          \and
        	  N. R. Walborn\inst{4}
        	  \and
	          A. Pellerin\inst{5}
	          \and
           S. Sim\'on-D{\'\i}az\inst{6,7}
	          \and
	          A. Sota\inst{8}
	          \and
	          A. Marco\inst{2}
	          \and
	          J. Alonso-Santiago\inst{2}
        	  \and
	          J. Sanchez Bermudez\inst{8}
        	  \and
        	  R. C. Gamen\inst{9}
        	  \and
        	  J. Lorenzo\inst{2}
          }

   \institute{Centro de Astrobiolog{\'\i}a, CSIC-INTA, campus ESAC, apartado postal 78, E-28\,691 Villanueva de la Ca\~nada, Madrid, Spain \\
	      \email{jmaiz@cab.inta-csic.es} \\
         \and
              Departamento de F{\'\i}sica, Ingenier{\'\i}a de Sistemas y Teor{\'\i}a de la Se\~nal, Escuela Polit\'ecnica Superior, Universidad de Alicante, P.O.~Box~99, E-03\,080 Alicante, Spain \\
         \and
              Departamento de F{\'\i}sica y Astronom{\'\i}a, Universidad de La Serena, Av. Cisternas 1200 Norte, La Serena, Chile \\
         \and
              Space Telescope Science Institute, 3700 San Martin Drive, Baltimore, MD 21\,218, USA \\
         \and
              Department of Physics \& Astronomy, 1 College Circle, SUNY Geneseo, Geneseo, NY 14\,454, USA \\
         \and
              Instituto de Astrof{\'\i}sica de Canarias, E-38\,200 La Laguna, Tenerife, Spain \\
         \and
              Departamento de Astrof{\'\i}sica, Universidad de La Laguna, E-38\,205 La Laguna, Tenerife, Spain \\
         \and
              Instituto de Astrof{\'\i}sica de Andaluc{\'\i}a-CSIC, Glorieta de la Astronom\'{\i}a s/n, E-18\,008 Granada, Spain \\
         \and
              Instituto de Astrof\'{\i}sica de La Plata (CCT La Plata-CONICET, Universidad Nacional de La Plata), Paseo del Bosque s/n, 1900 La Plata, Argentina \\
             }

   \date{Received XX XXX 2015; accepted XX XXX 2015}

 
  \abstract
  {It appears that most (if not all) massive stars are born in multiple systems. At the same time, the most massive binaries are hard to find due to their low numbers 
  throughout the Galaxy and the implied large distances and extinctions.}
  {We want to study: [a] \LSXI, identified in this paper as a very massive binary; [b] another nearby massive system, \LSXII; and [c] the surrounding stellar cluster, 
   \BXC.}
  {Most of the data used in this paper are multi-epoch high-S/N optical spectra though we also use Lucky Imaging and archival photometry. The spectra are reduced with devoted
   pipelines and processed with our own software, such as a spectroscopic-orbit code, CHORIZOS, and MGB.}
   {\LSXI\ is identified as a new very-early-O-type spectroscopic binary [O3.5 If* + O3.5 If*] and \LSXII\ as another early O-type system [O4.5 V((f))]. We measure a 97.2-day period  
   for \LSXI\ and derive minimum masses of 38.80$\pm$0.83~M$_\odot$ and 35.60$\pm$0.77~M$_\odot$ for its two stars. We measure the extinction to both stars, estimate the distance,
   search for optical companions, and study the surrounding cluster. In doing so, a variable extinction is found as well as discrepant results for the 
   distance. We discuss possible explanations and suggest that \LSXII\ may be a hidden binary system, where the companion is currently undetected.}
   {}


   \maketitle
%

\section{Introduction}

\begin{figure}
\centerline{\includegraphics[width=\linewidth]{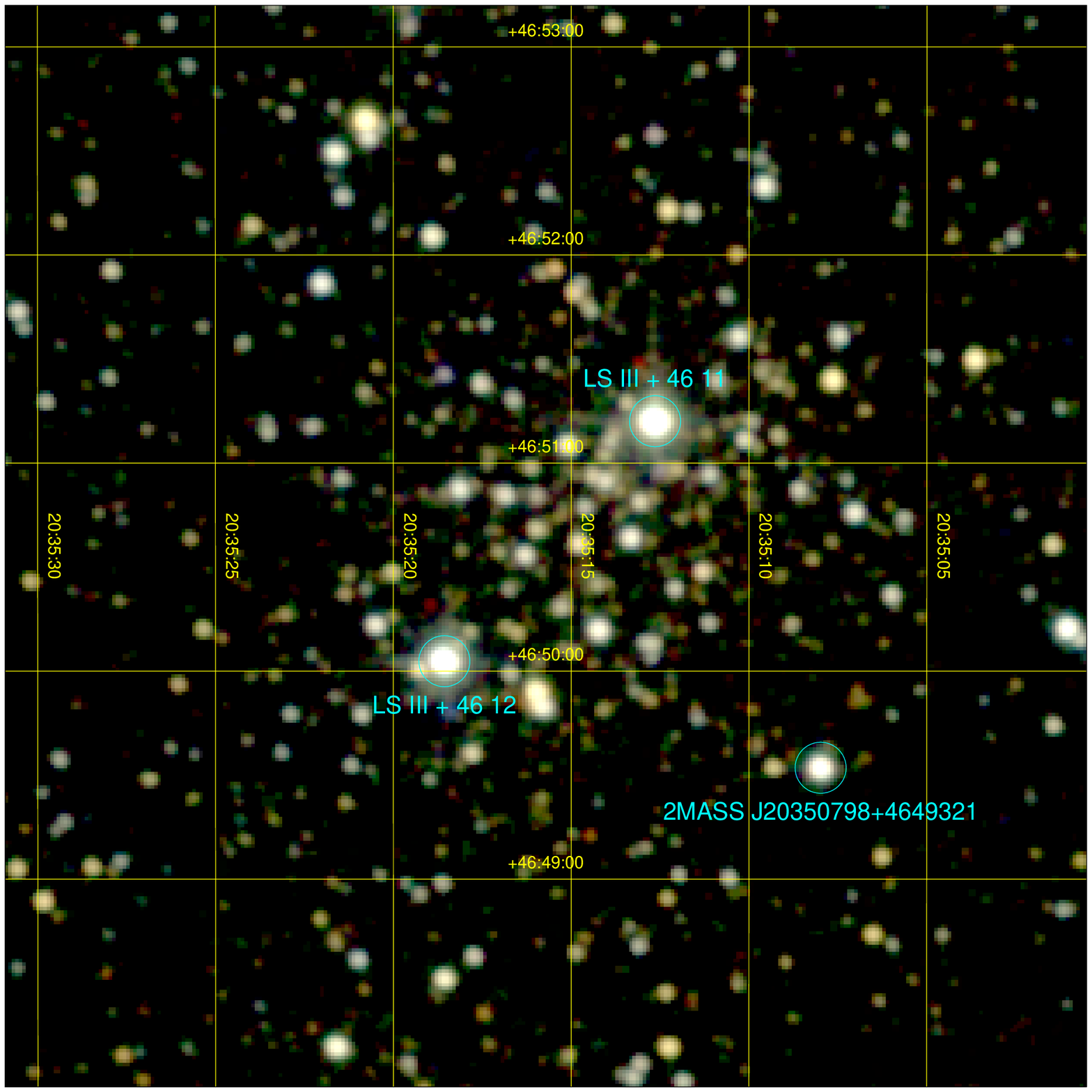}}
\caption{2MASS $K_{\mathrm{S}}HJ$ three-color RGB mosaic of \BXC. The intensity level in each channel is logarithmic.}
\label{2massimage}
\end{figure}

\begin{table*}
\centerline{
\begin{tabular}{lccl}
                & \LSXI\              & \LSXII\          & Reference                     \\
\hline
Sp. type        & O3.5~If*~+~O3.5~If* & O4.5~V((f))      & This work                     \\
RA (J2000)      & 20:35:12.642        & 20:35:18.566     & \citet{Hogetal00a}            \\
dec (J2000)     & +46:51:12.12        & +46:50:02.90     & \citet{Hogetal00a}            \\
$l$ (deg.)      & 84.8844             & 84.8791          & \citet{Hogetal00a}            \\
$b$ (deg.)      & +3.8086             & +3.7836          & \citet{Hogetal00a}            \\
$V_{J}$         & 10.889$\pm$0.021    & 10.268$\pm$0.009 & This work                     \\
$K_{\mathrm S}$ & 6.971$\pm$0.023     & 7.470$\pm$0.023  & \citet{Skruetal06}, this work \\
\hline
\end{tabular}
}
\caption{\LSXI\ and \LSXII\ summary.}
\label{summary}
\end{table*}

$\,\!$ \indent Multiplicity is an endemic disease among O stars \citep{Masoetal98,Sanaetal13b,Sotaetal14}. Indeed, it is difficult to find an O star that 
was born in isolation: among the \citet{Sotaetal14} sample of O7-O9.7 V-IV stars south of -20 deg, there is just one clear, currently single object, 
namely $\mu$~Col, which is however a known runaway and thus likely was born in a multiple system or compact cluster. 
Only a small fraction of the other types of O stars in the \citet{Sotaetal14} sample are apparently single but those cases are more 
luminous and can easily hide an e.g. main-sequence B-type companion in their glare. Both types of binarity, spectroscopic and visual, are common
among O stars and combinations of both types (implying higher-order multiplicities) are also relatively frequent, usually in the form of hierarchical
systems, with one pair in a close orbit and additional star(s) at larger separations. The OWN survey \citep{Barbetal10} has detected a peak around 10 days
in the spectroscopic period distribution of 240 southern massive stars.  Most of those systems will
interact during their evolution and many mergers are expected \citep{Sanaetal12a}. Even in those cases where interactions will not take place it is
important to characterize the binarity of O stars since ignoring it can lead to biases in the measured properties, such as derived masses and predicted 
colors of small stellar populations or young clusters.

The issue of multiplicity among massive stars also affects the measurement of the stellar upper mass limit, since one has to make sure that a very
luminous object is not in reality a combination of two or more objects, an issue that has produced a number of false alarms in the past, see e.g.
\citet{Maizetal07}. For a long time, all very massive stars were thought to be born as very early O-type objects but the most current information leans 
towards the extremely massive ones being born as WNh stars \citep{Crowetal10}. However, statistics are very poor to consider the case closed. For
example, only one O2 star with accurate blue-violet spectral classification is known in the Milky Way, HD~93\,129~AaAb \citep{Sotaetal14}, and it is
a multiple system located in the center of a dense stellar cluster. We clearly need to find more very massive stars to understand their properties
and evolution.

The sparse young open cluster \BXC\ (Fig.~\ref{2massimage}) is almost unstudied, but \citet{Sand74} suggested that the early-type stars \LSXI\ and \LSXII\ 
were members of the cluster. Both objects are classified as OB$^{+}$ in the Luminous Stars catalogue \citep{Hardetal64}, suggesting high luminosities. 
From an analysis of 2MASS 
data for the region, \citet{Tadr08} concluded that the cluster was young ($<$100~Ma old), located at 2.4~kpc, and reddened by $E(B-V)=1.15$~mag. 
\citet{Wend71} suggested that \LSXI\ and \LSXII\ were the ionizing sources for the \ion{H}{ii} region [GS55]~215, which can be identified with Sh~2-115
\citep{HartFell80}. \citet{MayeMaca73} derived a spectral type O6 and a distance of 2.3~kpc for \LSXII, but were unable to provide a spectral type for 
\LSXI, which is more reddened. \LSXI\ was identified as the counterpart of the moderately hard {\it ROSAT} source RX~J2035.2+4651 by \citet{Motcetal97},
during a search for possible high-mass X-ray binaries. They derived a spectral type O3-5~III(f)e from intermediate-resolution spectroscopy, and concluded
that \LSXI\ could produce the observed X-ray luminosity of $\sim 5\times10^{33}\:\mathrm{erg}\,\mathrm{s}^{-1}$ with ``probably no need'' 
for a compact companion but that ``the extreme value of the X-ray luminosity of this star would certainly deserve further detailed investigation'' without
mentioning an alternative (e.g. a colliding-wind binary). \citet{Motcetal97} also detected an X-ray source coincident with \LSXII\ with about one third of
the X-ray flux of \LSXI.

As part of our Galactic O-Star Spectroscopic Survey (GOSSS, \citealt{Maizetal11}), we observed \LSXI\ with the TWIN spectrograph of the 3.5~m 
telescope at Calar Alto in early November 2009. We immediately confirmed the very early nature of the system but more surprising was the double-lined
spectroscopic binary (SB2) nature
detected in the \ion{He}{ii} lines. Given the importance of the discovery, we observed the system on three consecutive nights (1+2+3 November) and we
noticed only a small decrease in the velocity separation, thus excluding a short period. For the next two years we reobserved \LSXI\ on different
occasions but we failed to detect clearly split \ion{He}{ii} lines until September 2011, when we observed it with the 4.2~m WHT at La Palma. The new
detection gave us a series of possible periods for the system and led to new observations in the subsequent years. By 2013 we had a preliminary orbit
with a period close to 97 days and by 2014 we obtained the final orbit, with a large eccentricity that explained the measurement difficulties.

In this paper we first describe the used data, both the primary spectroscopic observations and the complementary high-resolution imaging and archival
photometry. We then present our results: the spectral classification of \LSXI\ and \LSXII, the spectroscopic orbit of \LSXI, the extinction and 
distance to both stars, a search for visual companions, and a global analysis of \BXC. We finally discuss our results, including their relevance for 
the study of the stellar upper mass limit, and present our conclusions. In a subsequent paper we will analyze the foreground ISM in front of \BXC.

\section{Data}

\subsection{Spectroscopy}

\begin{table}
\tiny
\addtolength{\tabcolsep}{-3pt}
\begin{tabular}{cclc}
\hline
date         & HJD$-$2\,400\,000 & Telescope  & Wavelength range      \\
(YYYY-MM-DD) &                   &            & (\AA)                 \\
\hline
2009-11-01   & 55\,137.317       & CAHA-3.5 m & 3950-5050             \\
2009-11-02   & 55\,138.313       & CAHA-3.5 m & 3950-5050             \\
2009-11-03   & 55\,139.320       & CAHA-3.5 m & 3950-5050             \\
2009-11-03   & 55\,139.503       & CAHA-3.5 m & 3950-5050             \\
2009-12-05   & 55\,171.380       & OSN-1.5 m  & 5350-6750             \\
2009-12-07   & 55\,173.355       & OSN-1.5 m  & 5350-6750             \\
2009-12-09   & 55\,175.317       & OSN-1.5 m  & 5350-6750             \\
2009-12-10   & 55\,176.363       & OSN-1.5 m  & 5350-6750             \\
2010-05-01   & 55\,318.674       & OSN-1.5 m  & 5350-6750             \\
2010-06-17   & 55\,365.659       & OSN-1.5 m  & 5350-6750             \\
2010-06-26   & 55\,374.669       & OSN-1.5 m  & 5350-6750             \\
2011-06-13   & 55\,726.513       & WHT        & 3890-5570             \\
2011-06-15   & 55\,728.545       & WHT        & 3890-5570             \\
2011-06-21   & 55\,734.768       & HET        & 3811-4709 + 4758-5735 \\
2011-06-26   & 55\,739.779       & HET        & 5311-6275 + 6396-7325 \\
2011-09-09   & 55\,814.597       & WHT        & 3890-5570             \\
2011-09-10   & 55\,815.443       & WHT        & 3890-5570             \\
2011-09-11   & 55\,816.460       & WHT        & 3890-5570             \\
2011-09-12   & 55\,817.381       & WHT        & 3890-5570             \\
2011-09-12   & 55\,817.455       & NOT        & 3690-7180             \\
2011-09-12   & 55\,817.568       & WHT        & 3890-5570             \\
2011-09-13   & 55\,818.791       & HET        & 5311-6275 + 6396-7325 \\
2011-09-15   & 55\,820.489       & NOT        & 3690-7180             \\
2011-09-17   & 55\,822.365       & NOT        & 3690-7180             \\
2011-09-28   & 55\,833.751       & HET        & 3811-4709 + 4758-5735 \\
2011-10-02   & 55\,837.724       & HET        & 5311-6275 + 6396-7325 \\
2011-10-03   & 55\,838.724       & HET        & 5311-6275 + 6396-7325 \\
2011-10-06   & 55\,841.706       & HET        & 5311-6275 + 6396-7325 \\
2011-10-07   & 55\,842.706       & HET        & 5311-6275 + 6396-7325 \\
2011-10-10   & 55\,845.719       & HET        & 5311-6275 + 6396-7325 \\
2011-10-31   & 55\,866.649       & HET        & 5311-6275 + 6396-7325 \\
2011-11-15   & 55\,881.601       & HET        & 5311-6275 + 6396-7325 \\
2011-11-27   & 55\,893.590       & HET        & 5311-6275 + 6396-7325 \\
2012-04-11   & 56\,029.953       & HET        & 5311-6275 + 6396-7325 \\
2012-05-15   & 56\,063.877       & HET        & 5311-6275 + 6396-7325 \\
2012-05-28   & 56\,076.821       & HET        & 5311-6275 + 6396-7325 \\
2012-06-10   & 56\,089.815       & HET        & 5311-6275 + 6396-7325 \\
2012-07-06   & 56\,115.596       & WHT        & 3910-5590             \\
2012-07-08   & 56\,117.657       & WHT        & 3910-5590             \\
2012-08-27   & 56\,167.808       & HET        & 3811-4709 + 4758-5735 \\
2012-09-07   & 56\,178.537       & CAHA-2.2 m & 3925-9225             \\
2012-09-23   & 56\,194.424       & WHT        & 3910-5590             \\
2012-09-24   & 56\,195.442       & WHT        & 3910-5590             \\
2012-09-25   & 56\,196.480       & WHT        & 3910-5590             \\
2012-10-02   & 56\,203.713       & HET        & 5311-6275 + 6396-7325 \\
2012-10-04   & 56\,205.732       & HET        & 5311-6275 + 6396-7325 \\
2012-10-05   & 56\,206.727       & HET        & 5311-6275 + 6396-7325 \\
2012-10-22   & 56\,223.343       & CAHA-2.2 m & 3925-9225             \\
2012-12-04   & 56\,266.548       & HET        & 5311-6275 + 6396-7325 \\
2012-12-05   & 56\,267.377       & CAHA-2.2 m & 3925-9225             \\
2012-12-07   & 56\,269.531       & HET        & 5311-6275 + 6396-7325 \\
2012-12-30   & 56\,292.280       & CAHA-2.2 m & 3925-9225             \\
2013-06-14   & 56\,458.541       & WHT        & 3910-5590             \\
2013-09-30   & 56\,566.341       & CAHA-2.2 m & 3925-9225             \\
2013-10-15   & 56\,581.411       & WHT        & 3910-5590             \\
2013-10-16   & 56\,582.442       & WHT        & 3910-5590             \\
2013-10-17   & 56\,583.344       & WHT        & 3910-5590             \\
2014-04-13   & 56\,761.624       & CAHA-2.2 m & 3925-9225             \\
2014-04-14   & 56\,762.611       & CAHA-2.2 m & 3925-9225             \\
2014-06-07   & 56\,816.568       & CAHA-3.5 m & 3880-4980             \\
2014-07-12   & 56\,851.562       & CAHA-3.5 m & 3950-5050             \\
2014-08-04   & 56\,874.526       & NOT        & 3690-7180             \\
2014-08-05   & 56\,875.513       & CAHA-2.2 m & 3925-9225             \\
2014-08-05   & 56\,875.572       & NOT        & 3690-7180             \\
2014-08-06   & 56\,876.373       & CAHA-2.2 m & 3925-9225             \\
2014-08-07   & 56\,877.446       & CAHA-2.2 m & 3925-9225             \\
2014-08-08   & 56\,878.583       & NOT        & 3690-7180             \\
2014-08-14   & 56\,884.384       & CAHA-2.2 m & 3925-9225             \\
2014-08-16   & 56\,886.528       & CAHA-2.2 m & 3925-9225             \\
2014-08-18   & 56\,888.542       & CAHA-2.2 m & 3925-9225             \\
\hline
\end{tabular}
\addtolength{\tabcolsep}{3pt}
\normalsize
\caption{Spectroscopic observation log for \LSXI. The date refers to the local evening.}
\label{obs1}
\end{table}

\begin{table}
\tiny
\addtolength{\tabcolsep}{-3pt}
\begin{tabular}{cclc}
\hline
date         & HJD$-$2\,400\,000 & Telescope  & Wavelength range      \\
(YYYY-MM-DD) &                   &            & (\AA)                 \\
\hline
2011-06-13   & 55\,726.513       & WHT        & 3890-5570             \\
2011-06-15   & 55\,728.545       & WHT        & 3890-5570             \\
2011-09-09   & 55\,814.597       & WHT        & 3890-5570             \\
2011-09-10   & 55\,815.443       & WHT        & 3890-5570             \\
2011-09-11   & 55\,816.460       & WHT        & 3890-5570             \\
2011-09-12   & 55\,817.381       & WHT        & 3890-5570             \\
2011-09-12   & 55\,817.568       & WHT        & 3890-5570             \\
2012-07-06   & 56\,115.596       & WHT        & 3910-5590             \\
2012-07-08   & 56\,117.657       & WHT        & 3910-5590             \\
2012-08-15   & 56\,155.648       & HET        & 5311-6275 + 6396-7325 \\
2012-09-23   & 56\,194.424       & WHT        & 3910-5590             \\
2012-09-24   & 56\,195.442       & WHT        & 3910-5590             \\
2012-09-25   & 56\,196.480       & WHT        & 3910-5590             \\
2013-06-14   & 56\,458.541       & WHT        & 3910-5590             \\
2013-10-15   & 56\,581.411       & WHT        & 3910-5590             \\
2013-10-16   & 56\,582.442       & WHT        & 3910-5590             \\
2013-10-17   & 56\,583.344       & WHT        & 3910-5590             \\
2015-03-02   & 57\,084.715       & CAHA-2.2 m & 3925-9225             \\
2015-03-03   & 57\,085.717       & CAHA-2.2 m & 3925-9225             \\
2015-03-28   & 57\,110.696       & CAHA-2.2 m & 3925-9225             \\
\hline
\end{tabular}
\addtolength{\tabcolsep}{3pt}
\normalsize
\caption{Spectroscopic observation log for \LSXII. The date refers to the local evening.}
\label{obs2}
\end{table}

$\,\!$ \indent The spectroscopic data for \LSXI\ used in this paper were obtained with six different telescopes and are listed in 
Table~\ref{obs1}. Here we describe the data grouped under the different projects in which they were obtained.

{\it 1. GOSSS}: the Galactic O-Star Spectroscopic Survey is described in \citet{Maizetal11}. The GOSSS spectra used here were obtained with the TWIN 
spectrograph of the 3.5~m telescope at Calar Alto (CAHA) and the ISIS spectrograph of the 4.2~m William Herschel Telescope (WHT) at La Palma. The spectral 
resolving power was measured in each case\footnote{Note that the GOSSS spectra shown in \citet{Sotaetal11a,Sotaetal14} have been homogenized to $R = 2500$ 
whenever the resolving power was slightly higher.} and determined to be between 2800 and 3400 (TWIN) and between 3900 and 5400 (ISIS). The typical S/N 
per resolution element of the data is 300. The spectral range of the WHT data includes both \HeII{4541.591} and \HeII{5411.53} while that of the CAHA-3.5~m 
data included only the former\footnote{Throughout this work we refer to each line with the exact rest wavelength we are using in order to avoid problems with
future comparisons of absolute velocity values. The different number of significant digits reflects the uncertainty in our knowledge of the rest wavelengths 
and/or the single/multiplet nature}. GOSSS is also obtaining spectroscopy with the Albireo spectrograph of the 1.5~m 
telescope at the Observatorio de Sierra Nevada (OSN) but its aperture is too small to obtain a good S/N in the blue-violet. However, we also obtained some 
OSN yellow-red spectra at $R\sim 2000$ in order to study \HeII{5411.53}. 

{\it 2. NoMaDS}: the Northern Massive Dim Survey is described in \citet{Maizetal12} and \citet{Pelletal12}. NoMaDs spectra were obtained with the High
Resolution Spectrograph of the 9-m Hobby-Eberly Telescope (HET) at McDonald Observatory. Their spectral resolving power is 30\,000 and all epochs include 
\HeII{5411.53} though two different setups were used on different occasions: one that goes from the violet to the yellow with a small gap (3811-4709~\AA\ + 
4758-5735~\AA) and one that goes from the green to the red with another small gap (5311-6275~\AA\ + 6396-7325~\AA).

{\it 3. IACOB}: the Instituto de Astrof{\'\i}sica de Canarias OB database is described in \citet{SimDetal11c} and \citet{SimDetal15b}. IACOB spectra were 
obtained with the FIES spectrograph at the 2.6~m Nordic Optical Telescope (NOT) at la Palma. Their spectral resolving power is 23\,000 and they cover the 
whole optical range. Note that some epochs were obtained as part of IACOB itself and some were obtained during service time after a request to observe \LSXI\ 
was granted additional time.

{\it 4. CAF\'E-BEANS}: the Calar Alto Fiber-fed \'Echelle Binary Evolution Andalusian Northern Survey is described in \citet{Neguetal14}. CAF\'E-BEANS spectra 
were obtained with the CAF\'E spectrograph at the 2.2~m telescope at Calar Alto. Their spectral resolving power is 65\,000 and they cover the whole optical 
range.

The spectroscopic data for \LSXII\ were obtained with GOSSS, NoMaDS, and CAF\'E-BEANS and are listed in Table~\ref{obs2}. Note that the GOSSS instruments are long-slit
spectrographs so the \LSXI\ and \LSXII\ spectra were obtained simultaneously.

The GOSSS spectroscopy was reduced with the pipeline described by \citet{SotaMaiz11} while the NoMaDS, IACOB, and CAF\'E-BEANS were reduced with specific
pipelines for each instrument. The telluric lines in the high-resolution spectra were eliminated according to the procedure of \citet{Gardetal13}. Note that the 
high-resolution spectra cover a large range in wavelength and that \LSXI\ and \LSXII\ are moderately reddened. That means that the S/N is highly variable as a function 
of wavelength and that in the violet extreme only the NoMaDS spectra have a good S/N.

\subsection{High-resolution imaging}

$\,\!$ \indent To complement the spectroscopic observations of \LSXI\ and \LSXII\ we also present high-resolution imaging data that are part of a visual 
multiplicity survey of massive stars we are conducting using the Lucky Imaging instruments AstraLux Norte (at the 2.2~m CAHA telescope) and AstraLux Sur 
(at the 3.6~m NTT telescope at La Silla). The Lucky Imaging technique involves taking a large number (typically 10\,000) of very short exposures (typically, 30 ms),
selecting a small fraction of those based on the PSF quality, and then combining them with a drizzle-type algorithm.
The survey itself was presented in \citet{Maiz10a}; the reader is referred to that paper for details on the
data and processing. Later AstraLux results for individual stars are discussed in \citet{Sotaetal14} and \citet{SimDetal15a}. \citet{Maiz10a} used a 
Sloan-$z$-band image of \LSXI. Here we use that image plus two additional $z$-band AstraLux Norte images of \LSXI, two Sloan-$i$-band images of \LSXI, and 
one $z$-band image of \LSXII, obtained between 2007 and 2013.

\subsection{Archival photometry}

$\,\!$ \indent We used Simbad and Vizier to compile photometric data (Johnson $UBV$, Tycho-2 $BV$, and 2MASS $JHK_{\mathrm{S}}$) from the literature for \LSXI\ 
and \LSXII. The photometry was processed with CHORIZOS, as explained in the next section.  

We also compiled the IPHAS and 2MASS photometry of the \BXC\ and foreground populations. For the case of \LSXI, the 2MASS PSC gives $J$ and $H$
magnitudes but not a $K_{\mathrm{S}}$ value, where an X flag appears, indicating that ``'there is a detection at this location, but no valid brightness estimate 
can be extracted using any algorithm''. For \LSXII, which has similar NIR magnitudes (see Table~\ref{twomass}), the 2MASS PSC gives the three 
magnitudes with good quality flags. We downloaded the 2MASS images to look into this issue.

\LSXI\ and \LSXII\ have NIR magnitudes that make them slightly saturated (by $\sim$1 magnitude) in the 2MASS images (which have saturation levels of 9.0, 8.5, and 
8.0 for $J$, $H$, and $K_{\mathrm{S}}$, respectively). For that reason, their 2MASS PSC magnitudes were obtained by aperture photometry in the 51~ms ``Read\_1''
exposures which, unfortunately, are not publicly available. However, for such a small level of saturation only 1-4 pixels (depending on the star centering) are 
expected to be saturated so it should be possible to do differential (between the two stars) aperture photometry with an inner and an outer integration radius. 
We tested this on \LSXI\ 
and \LSXII\ using an inner radius of 1.5~pixels and an outer radius of 4.0~pixels and found $\Delta J$~=~0.304 and $\Delta H$~=~0.412, values that are within
one sigma of those in the 2MASS PSC (0.281$\pm$0.031 and 0.424$\pm$0.025, respectively, see Table~\ref{twomass}), lending credibility to the technique. We
looked at the \LSXI\ $K_{\mathrm{S}}$ radial profile and noticed no peculiarities so we applied the same technique, which yielded $\Delta K_{\mathrm{S}}$~=~0.499.
Therefore, we derive a value of $K_{\mathrm{S}} = 6.971\pm 0.023$ for \LSXI, where we adopted as uncertainty the same value as for \LSXII.

\section{Results}

\subsection{Spectral classification of \LSXI\ and \LSXII}

\begin{figure*}
\centerline{\includegraphics[width=\linewidth, bb=53 28 546 351]{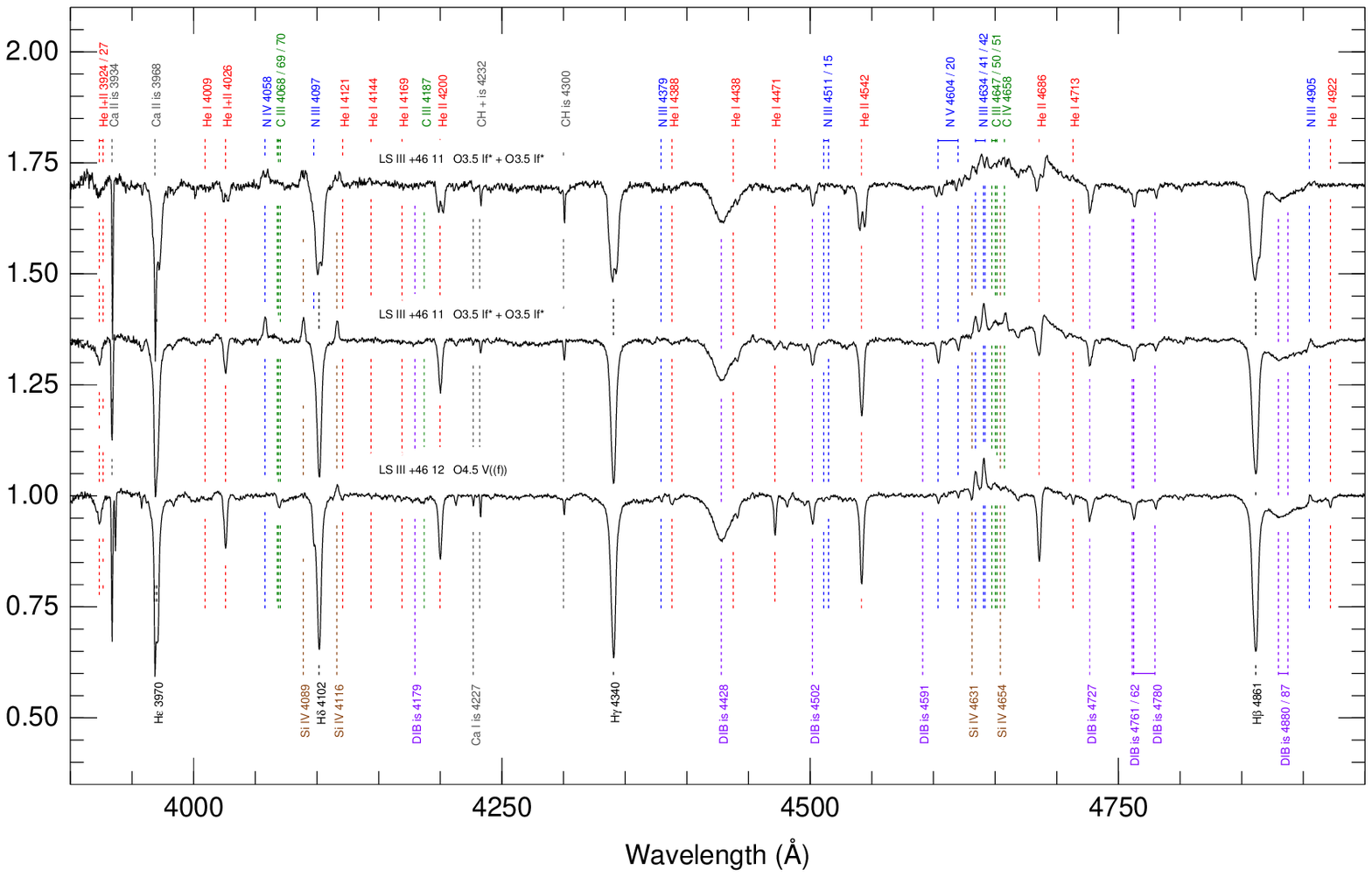}}
\caption{Sample GOSSS spectra of \LSXI\ and \LSXII\ in the classical blue-violet spectral classification range. The top spectrum shows an example of \LSXI\ near 
maximum velocity separation, the middle one an example of \LSXI\ near minimum velocity separation, and the bottom one an example of \LSXII. The three cases show 
WHT data with the original spectral resolving power.}
\label{gosss_spect}
\end{figure*}

\begin{figure}
\centerline{\includegraphics[width=\linewidth]{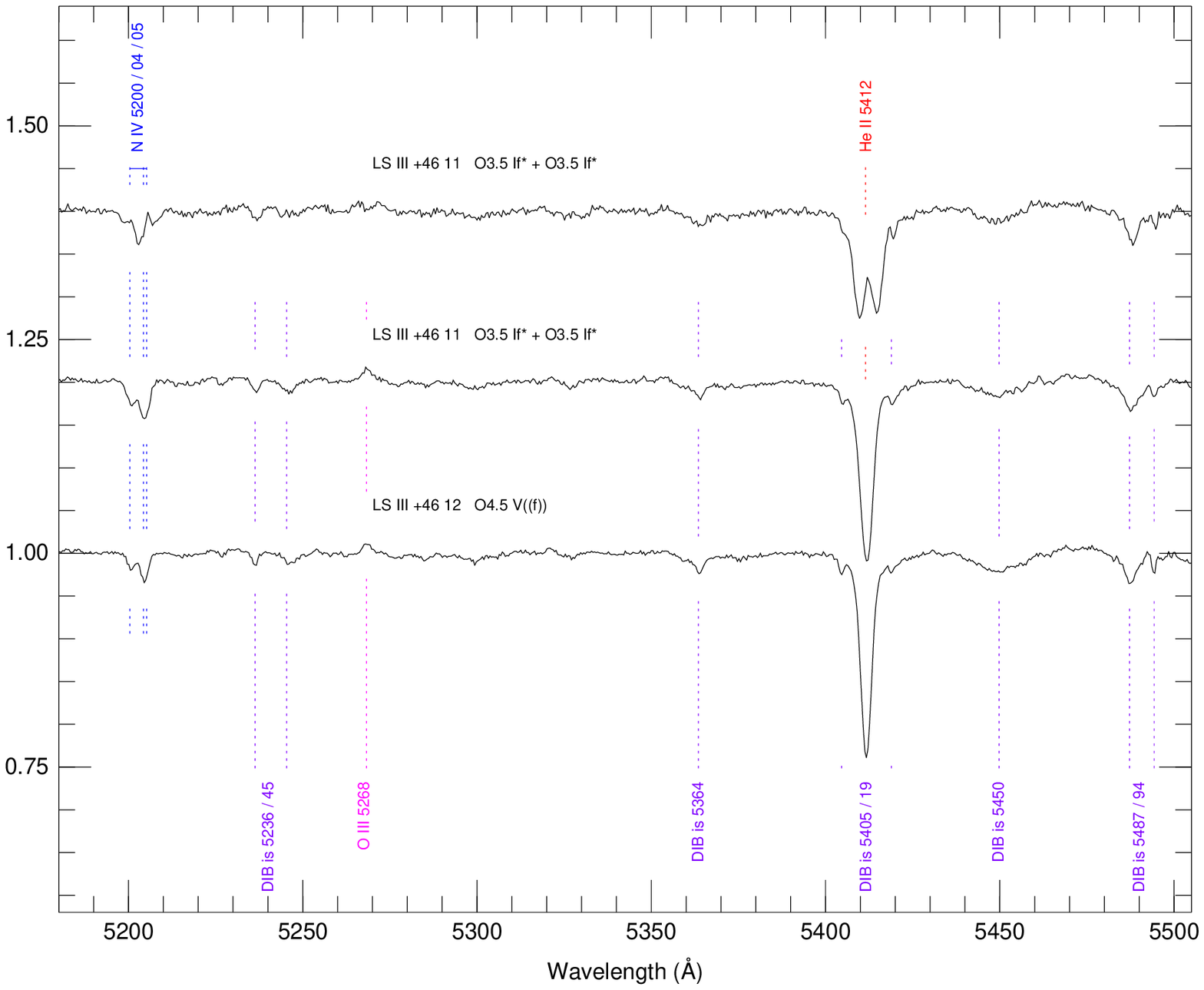}}
\caption{Same as Fig.~\ref{gosss_spect} for the wavelength range around \HeII{5411.53}, the primary line we use for the stellar velocity measurements.}
\label{gosss_spect2}
\end{figure}

$\,\!$ \indent The top two plots of Figures~\ref{gosss_spect}~and~\ref{gosss_spect2} show two GOSSS spectra of \LSXI, one near maximum velocity separation and 
one close to the point where both stars have the same velocity. In the first plot, the spectra of the two components are virtually 
indistinguishable, with all the relevant stellar lines showing very similar ratios and widths. The only difference can be seen in the 
different absolute depths, the ones of the primary being $\sim10\%$ more intense than those of the secondary. Therefore, we assign
the same spectral type to both components.

We classified the spectra using MGB \citep{Maizetal12} and v2.0 of the GOSSS standard grid \citep{Maizetal14c}. In the middle plot of
Figure~\ref{gosss_spect}, \HeI{4471.507} appears as a very weak line in comparison with \HeII{4541.59} and \NIV{4057.75} is in emission with a 
similar intensity as \NIII{4640.64+4641.85}, indicating a spectral type of O3.5 \citep{Walbetal02b}. \HeII{4685.71} has a P-Cygni 
profile\footnote{Note that the absorption component is only slightly blueshifted.} (which
appears to be present in both components, see the top plot) yielding a luminosity class of I since II is not defined at O3.5.
Therefore, given the required f suffix (Table 2 in \citealt{Sotaetal14}), \LSXI\ is an O3.5~If*~+~O3.5~If*.

The bottom plot of Figure~\ref{gosss_spect} shows a GOSSS spectrum of \LSXII. All the GOSSS spectra are consistent with a constant
spectral type i.e. we detect no sign of the target being an SB2. The GOSSS spectra are not calibrated in absolute velocity (the spectra
are left in the star reference frame) but a comparison with prominent ISM lines shows no sign of relative velocity shifts, so we do not
detect an SB1 character for \LSXII, either. 

\NIV{4057.75} is not seen in emission in the spectrum of \LSXII\ and \HeI{4471.507} is significantly stronger than in \LSXI, yielding a spectral
type of O4.5. \HeII{4685.71} is quite deep, so the luminosity class is V. Since \NIII{4640.64+4641.85} is clearly in emission, \CIII{4647.419+4650.246+4651.473} is
weak, and there are no signs of line broadening, the final spectral type is O4.5~V((f)).

Note that \NIV{5200.41+5204.28+5205.15} is seen in absorption in all cases, as expected for these spectral types \citep{GameNiem02}.

\subsection{The spectroscopic orbit of \LSXI}

\begin{table}
\centerline{
\addtolength{\tabcolsep}{-5pt}
\begin{tabular}{lr@{.}lcr@{.}ll}
\hline
$P$                &     97&193 &$\pm$&0&010        & days                 \\
$T_0$              &56\,002&504 &$\pm$&0&088        & HJD$-$2\,400\,000    \\
$K_{12}$           &    237&3   &$\pm$&1&2          & km/s                 \\
$e$                &      0&5685&$\pm$&0&0036       &                      \\
$\omega$           &    124&89  &$\pm$&0&62         & degrees              \\
$a_{12}\sin i$     &      1&7438&$\pm$&0&0077       & AU                   \\
$(M_1+M_2)\sin^3i$ &     74&9   &$\pm$&1&0          & M$_\odot$            \\
$n_{\rm epochs}$   & \multicolumn{2}{c}{70$\;\;\,$} & \multicolumn{4}{c}{} \\
$n_{\rm dof}$      & \multicolumn{2}{c}{65$\;\;\,$} & \multicolumn{4}{c}{} \\
$\chi^2_{\rm red}$ &      1&43                      & \multicolumn{4}{c}{} \\
\hline
\end{tabular}
\addtolength{\tabcolsep}{5pt}
}
\caption{\LSXI\ spectroscopic binary results for $\Delta v \equiv v_2-v_1$. $K_{12}$ is the radial velocity amplitude of the relative orbit.}
\label{SB1}
\end{table}

\begin{table}
\centerline{
\addtolength{\tabcolsep}{-5pt}
\begin{tabular}{lr@{.}lcr@{.}ll}
\hline
$P$                        &     97&168 &$\pm$&0&025        & days                 \\
$T_0$                      &56\,002&80  &$\pm$&0&25         & HJD$-$2\,400\,000    \\
$K_1$                      &    112&7   &$\pm$&1&2          & km/s                 \\
$K_2$                      &    122&9   &$\pm$&1&3          & km/s                 \\
$e$                        &      0&5627&$\pm$&0&0061       &                      \\
$\omega$                   &    126&1   &$\pm$&1&2          & degrees              \\
$\gamma_1$                 &    -17&68  &$\pm$&0&98         & km/s                 \\
$\gamma_2$                 &    -20&88  &$\pm$&1&04         & km/s                 \\
$a_1\sin i$                &      0&8325&$\pm$&0&0078       & AU                   \\
$a_2\sin i$                &      0&9073&$\pm$&0&0081       & AU                   \\
$M_1^3\sin^3i/(M_1+M_2)^2$ &      8&15  &$\pm$&0&98         & M$_\odot$            \\
$M_2^3\sin^3i/(M_1+M_2)^2$ &     10&55  &$\pm$&0&28         & M$_\odot$            \\
$M_2/M_1$                  &      0&9175&$\pm$&0&0095       &                      \\
$M_1\sin^3i$               &     38&80  &$\pm$&0&83         & M$_\odot$            \\
$M_2\sin^3i$               &     35&60  &$\pm$&0&77         & M$_\odot$            \\
$n_{\rm epochs}$           & \multicolumn{2}{c}{44$\;\;\,$} & \multicolumn{4}{c}{} \\
$n_{\rm dof}$              & \multicolumn{2}{c}{80$\;\;\,$} & \multicolumn{4}{c}{} \\
$\chi^2_{\rm red}$         &      0&95                      & \multicolumn{4}{c}{} \\
\hline
\end{tabular}
\addtolength{\tabcolsep}{5pt}
}
\caption{\LSXI\ spectroscopic binary results for $v_1$ + $v_2$. $K_1$ and $K_2$ are the radial velocity amplitudes of each orbit.}
\label{SB2}
\end{table}

$\,\!$\indent Most spectroscopic orbits for OB stars are studied with He\,{\sc i} lines (or even with metallic lines for B stars)
because they are intrinsically narrower than He\,{\sc ii} lines. However, for a system composed of two O3.5 stars He\,{\sc i} lines are
not practical because they are too weak and we are forced to resort to the broader He\,{\sc ii} lines\footnote{This is a factor that
likely contributed to the previous failure of the detection of the SB2 character of \LSXI.}. The strongest He\,{\sc ii} optical 
absorption lines in an O3.5~If* are \HeII{4541.591} and \HeII{5411.53}. The large extinction experienced by \LSXI\ (see below) makes a given S/N
easier to attain with the latter than with the former, so we selected \HeII{5411.53} as our primary line. Note, however, that the 
CAHA-3.5~m spectra do not include it. In those cases we used \HeII{4541.591} instead to measure the $\Delta v\equiv v_2-v_1$, the velocity difference
between the primary and the secondary. Also note that the CAHA-3.5~m and OSN-1.5~m spectra were not calibrated in absolute velocity and that the WHT 
spectra were calibrated using ISM lines (instead of the lamps) present in both the WHT and the high-resolution spectra (which were calibrated using
lamps). 

All the epochs were fitted simultaneously with an IDL code by leaving the intensity and width of each component fixed (allowing for the 
different resolutions of each spectrograph) and fitting a four-component Gaussian for \HeII{5411.53} (the two stellar components plus the
two DIBs at 5404.56~\AA\ and 5418.87~\AA, with the DIBs fixed at the same velocity for all epochs) 
and a two-component Gaussian for \HeII{4541.591}. The initial period search was conducted using an 
IDL implementation of the information entropy algorithm of \citet{Cincetal95}. The orbit fitting itself (including the final period 
calculation) was done independently by two of us: [a] J.M.A. using a code developed by himself with the help of a previous routine written by 
R.C.G. and [b] R.H.B. using an improved version of the \citet{BertGrob69} code. Results were compared and found to be compatible, so only the 
first ones will be reported here.

\begin{figure*}
\centerline{\includegraphics[width=0.490\linewidth, bb=28 28 566 566]{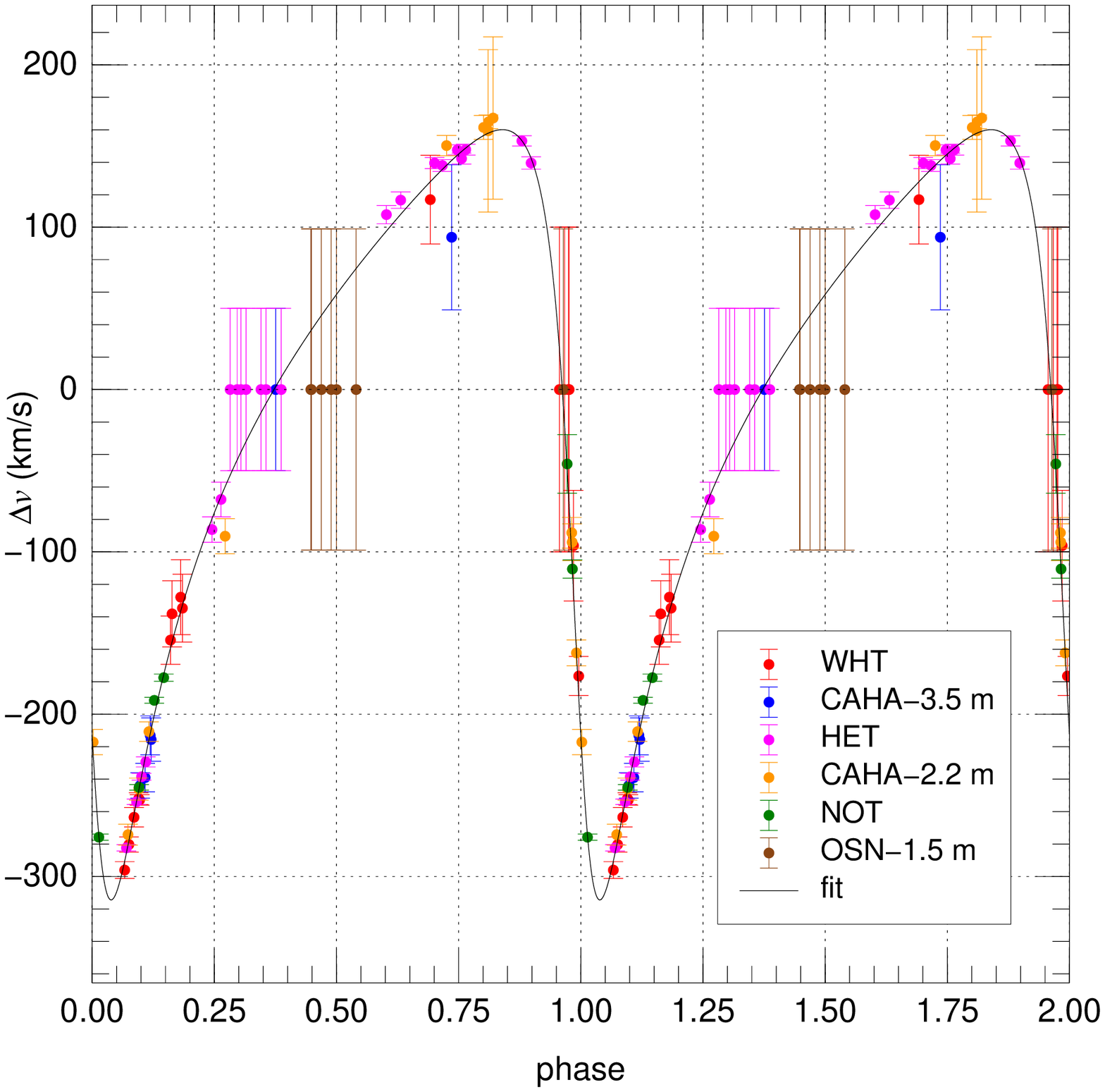} \
            \includegraphics[width=0.490\linewidth, bb=28 28 566 566]{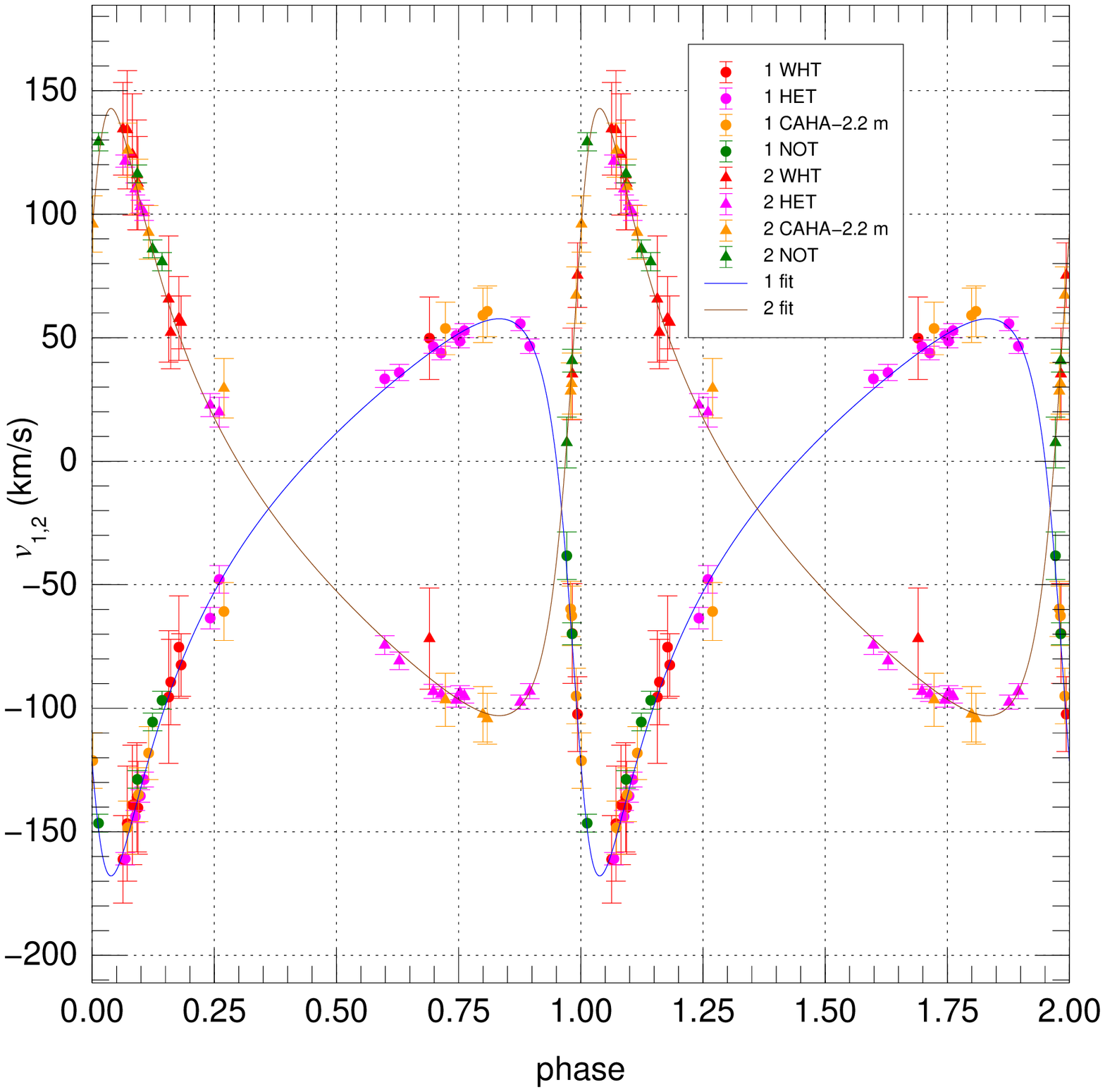}}
\caption{Phased radial velocity curves for $\Delta v$ (left) and $v_1$ + $v_2$ (right). The color code identifies the telescope. Note that the left
plot includes all data points while the right plot excludes the two telescopes without accurate absolute velocity calibration (CAHA-3.5 m and 
OSN-1.5 m).}
\label{RV}             
\end{figure*}

A first analysis was carried out with $\Delta v\equiv v_2-v_1$ and all the observations in Table~\ref{obs1}. Results are displayed in
Table~\ref{SB1} and plotted in the left panel of Figure~\ref{RV}. In some cases with $\Delta v$ close to zero it is not possible to
distinguish which component was the one with the larger velocity, since the two components are blended into a single Gaussian
and, given the similar fluxes, switching them around does not appreciably change the reduced $\chi^2$ or \chir\ of the fit. For those
epochs with identification confusion we used a $\Delta v$ of zero with large error bars. The orbit obtained with $\Delta v$ has a good 
fit (\chir = 1.43), a period consistent with the initial estimate, a relatively large eccentricity (as expected), and a very large
minimum system mass (uncorrected for inclination) of 74.9~M$_\odot$.

A second analysis was carried out with the separate $v_1$ and $v_2$ measurements. For this second analysis we excluded the CAHA-3.5~m
spectra (for which we only had \HeII{4541.591} and were not calibrated in absolute velocity), the OSN-1.5~m spectra (also uncalibrated in
absolute velocity and with identification confusion in all cases, since we always caught \LSXI\ near $\Delta v = 0$ with that
configuration), and the WHT spectra with identification confusion. Results are displayed in Table~\ref{SB2} and plotted in the right panel
of Figure~\ref{RV}. The value of \chir\ is even better than for the first analysis and the results are overall compatible. The
uncertainties of $P$, $T_0$, $e$, and $\omega$ are slightly larger due to the exclusion of some epochs. The mass ratio is close to the
intensity ratio of the \HeII{5411.53} or \HeII{4541.59} lines and in the same sense: the component with the stronger lines is the most massive.

The values of $\gamma_1$ and $\gamma_2$ in Table~\ref{SB2} indicate that the \LSXI\ center-of-mass radial velocity is in the range between -21~km/s and -17~km/s 
using the \HeII{5411.53} line. Fitting Gaussians to the \LSXI\ high-resolution spectra with better S/N and lowest velocity separation also yield values in that 
range.  However, doing the same with other lines gives different values: -27$\pm$2~km/s for \HeI{5875.65}, -23$\pm$4~km/s for \OIII{5592.25}, and -11$\pm$3~km/s 
for \HeII{8236.79}. These discrepancies are not uncommon when analyzing early-type O stars due to the intrinsic width of their lines and the effect of winds. 
We can conclude that the true center-of-mass radial velocity of \LSXI\ is between -25~km/s and -15~km/s without being able to provide a more precise measurement 
at this time.

We also analyzed the velocity of \LSXII\ using the four high-resolution epochs listed in Table~\ref{obs2}. We were unable to detect clear radial velocity 
variations at a level of 10 km/s or higher but with such a small number of epochs it is not possible to ascertain the spectroscopic binarity of the system.
We measured the radial velocity of the system using the same four spectral lines as for \LSXI\ and obtained a smaller range between -16~km/s and -10~km/s. The 
better agreement between lines is possibly caused by the weaker winds in the dwarf compared to the supergiant pair. These results are consistent with \LSXII\ 
being a spectroscopic single with the same center-of-mass radial velocity as \LSXI\, as the stronger winds of the latter likely bias its measured velocity towards
more negative (blueshifted) values. Also, the experience with \LSXI\ and other systems indicates that we should not exclude the possibility of \LSXII\ being a 
spectroscopic system with a period of months or longer.

\subsection{CHORIZOS analysis of \LSXI\ and \LSXII}

\begin{figure*}
\centerline{\includegraphics[width=0.482\linewidth, bb=28 28 566 566]{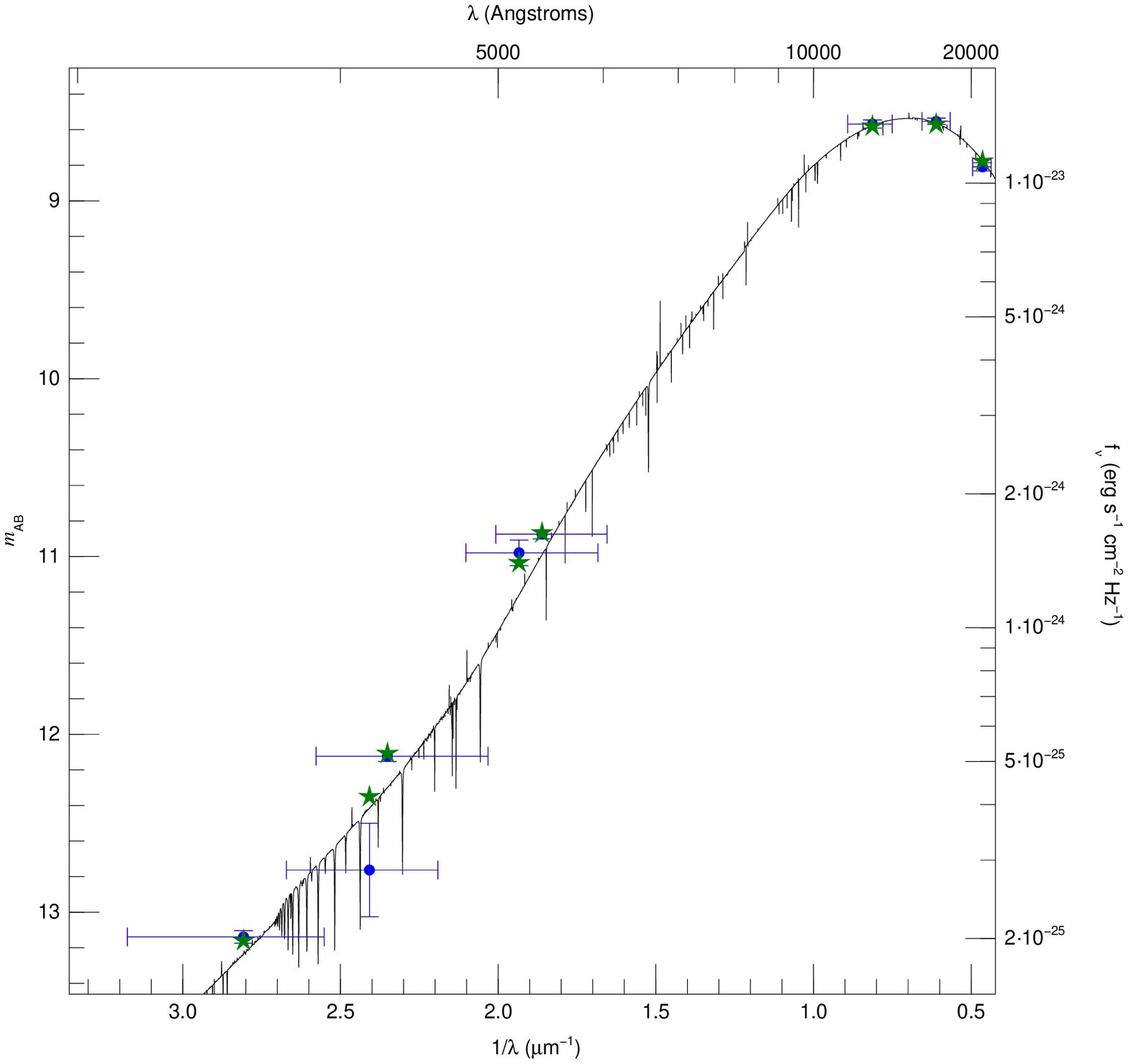} \
            \includegraphics[width=0.498\linewidth, bb=28 35 566 566]{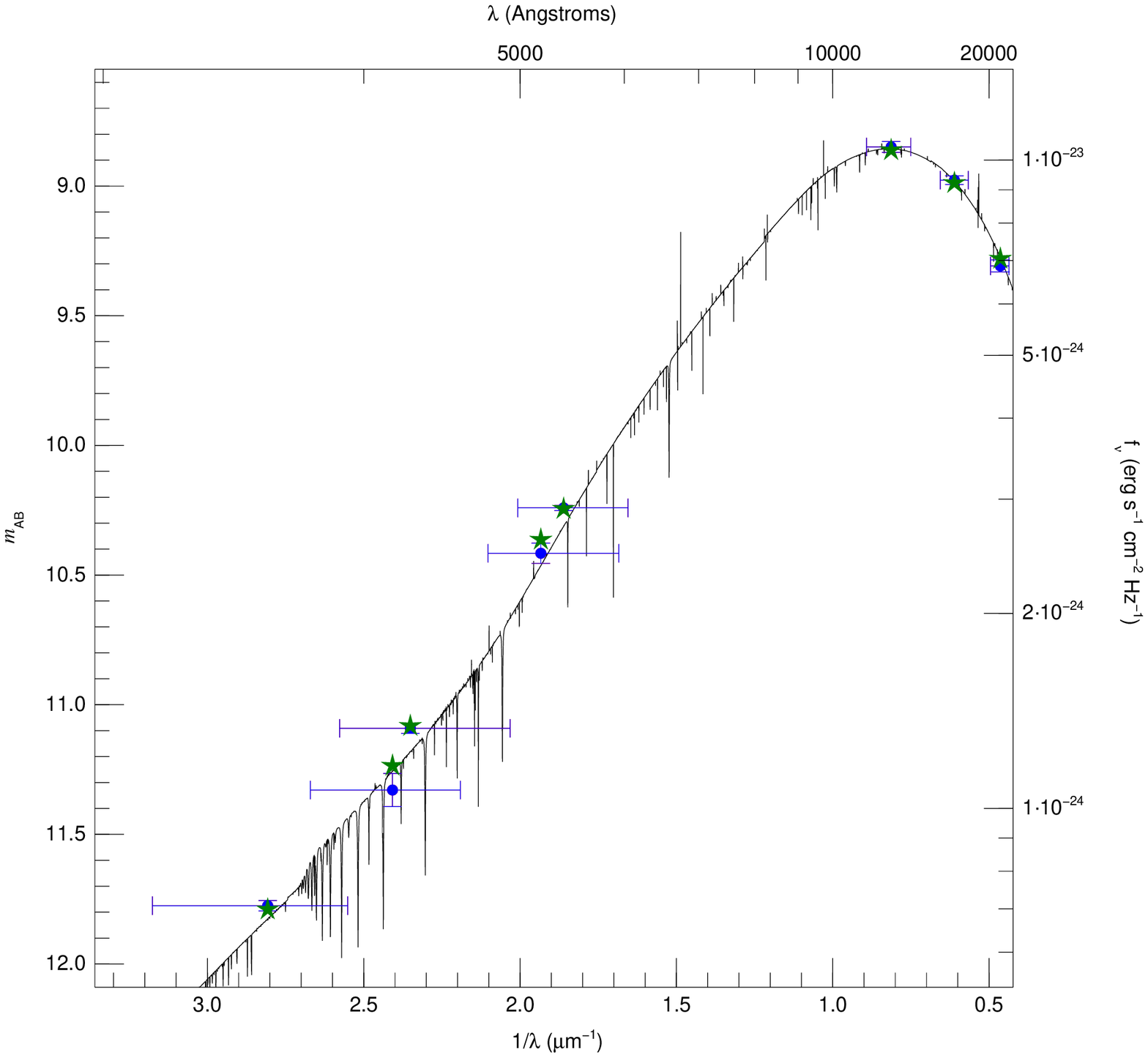}}
\caption{Best SED CHORIZOS fits for \LSXI\ (left, luminosity class of 1.0 used) and \LSXII\ (right, luminosity class of 5.0 used). Blue data points 
are used for the input photometry (vertical error bars indicate photometric uncertainties, horizontal ones approximate filter extent) and green stars 
for the synthetic photometry.}
\label{chorizos_spect}
\end{figure*}

\begin{table}
\begin{tabular}{lr@{$\pm$}lr@{$\pm$}l}
Quantity           & \multicolumn{2}{c}{\LSXI}   & \multicolumn{2}{c}{\LSXII}  \\
\hline
\teff (K)          & \multicolumn{2}{c}{41\,300} & \multicolumn{2}{c}{41\,900} \\
luminosity class   & \multicolumn{2}{c}{1.0}     & \multicolumn{2}{c}{5.0}     \\
\chir              & \multicolumn{2}{c}{1.36}    & \multicolumn{2}{c}{1.42}    \\
\rv                &  3.303  & 0.058             &  3.377  & 0.040             \\
\ebv\ (mag)        &  1.653  & 0.020             &  1.255  & 0.011             \\
\AV\  (mag)        &  5.475  & 0.037             &  4.272  & 0.021             \\
$V_{J,0}$ (mag)    &  5.414  & 0.021             &  5.995  & 0.017             \\
\logd\ (pc, min)   & \multicolumn{2}{c}{3.308}   & \multicolumn{2}{c}{3.103}   \\
\logd\ (pc, used)  & \multicolumn{2}{c}{3.407}   & \multicolumn{2}{c}{3.172}   \\
\logd\ (pc, max)   & \multicolumn{2}{c}{3.525}   & \multicolumn{2}{c}{3.276}   \\
\hline
\end{tabular}
\caption{Results of the CHORIZOS fits for \LSXI\ and \LSXII.}
\label{chorizos_output}
\end{table}

$\,\!$\indent We processed the Johnson+Tycho+2MASS photometry of \LSXI\ and \LSXII\ using the latest version of the CHORIZOS code \citep{Maiz04c}  to 
determine the amount and type of extinction and the distance to the stars. The CHORIZOS runs were executed with the following conditions:

\begin{itemize}
 \item We used the Milky Way grid of \citet{Maiz13a}, in which the two grid parameters are effective temperature (\teff) and photometric luminosity class 
       (LC). The latter quantity is defined in an analogous way to the spectroscopic equivalent but instead of being discrete it is a 
       continuous variable that varies from 0.0 (highest luminosity for that \teff) to 5.5 (lowest luminosity for that \teff). Note that the range is 
       selected in order to make objects with spectroscopic luminosity class V (dwarfs) have LC$\approx$5 and objects with spectral luminosity class I
       have LC$\approx$1. For O stars the spectral energy distributions (SEDs) are TLUSTY \citep{LanzHube03}.
 \item The extinction laws were those of \citet{Maizetal14a}, which are a single-family parameter with the type of extinction defined by \rv. The 
       amount of extinction is parameterized by \ebv. See \citet{Maiz13b} for their relationship with $R_V$ and $E(B-V)$ and why those quantities are 
       not good choices to characterize extinction.
 \item The \teff-spectral type conversion used is an adapted version of \citet{Martetal05a} that includes the spectral subtypes used by
       \citet{Sotaetal11a,Sotaetal14}. 
 \item \teff\ and LC were fixed while \rv, \ebv, and \logd\ were left as free parameters. The values of the \teff\ were established from the used
       \teff-spectral type conversion (see previous point). For LC we explored a range of possible values by doing CHORIZOS runs with 0.0, 0.5, 1.0, 1.5, and
       2.0 for \LSXI\ and runs with 4.0, 4.5, 5.0, and 5.5 for \LSXII. 
\end{itemize}

The CHORIZOS results are shown in Table~\ref{chorizos_output} and the best SEDs are plotted in Figure~\ref{chorizos_spect} along with the input and synthetic
photometry. 

\begin{itemize}
 \item As expected, the different runs for a given star with different values of LC give nearly identical results for \chir, \rv, and \ebv\ and only differ
       in \logd. This happens because the optical+NIR colors of O stars are nearly independent of luminosity\footnote{The largest difference takes place in
       the $K_{\mathrm{S}}$ band due to the wind contribution but that effect is not taken into account in the TLUSTY SEDs and is only expected to be of the 
       order of 0.01 magnitudes for the cases of interest here.} for a fixed \teff. Therefore, here we will concentrate on the results for LC=1.0 (\LSXI) and
       LC=5.0 (\LSXII) and consider the other runs only when discussing the distance.
 \item The values of \chir\ indicate that the fit is good, even for two cases such as these where the extinction is considerable. We also ran alternative 
       CHORIZOS executions using the \citet{Cardetal89} and \citet{Fitz99} extinction laws. For \citet{Cardetal89} the \chir\ were similar, as expected 
       for stars with moderate extinction with broad-band photometry and \rv\ values close to the canonical 3.1. For \citet{Fitz99} the \chir\ were significantly
       worse (by a factor of $\approx$2). Therefore, these results are another sign of the validity of the \citet{Maizetal14a} extinction laws for Galactic targets.
 \item The two stars show values of \rv\ which are compatible between them and only slightly larger than the canonical value of 3.1. Therefore, the same type 
       of dust appears to lie between each star and us and the grain size is typical for the Milky Way.
 \item The \LSXI\ extinction is significantly higher than that of \LSXII, which explains the similar magnitudes of the two objects despite \LSXI\ being expected 
       to be intrinsically more luminous and located at the same distance. This result also implies that there is significant differential extinction within
       the \BXC\ field. The value of \ebv\ for \LSXII\ is close to the $E(B-V)~=~1.15$ result for \BXC\ of \citet{Tadr08}. 
 \item The values for \logd\ listed are the uncorrected CHORIZOS output and are equivalent to spectroscopic parallaxes. They do not take into account the
       fact that \LSXI\ is an SB2 system with two components with similar luminosities while \LSXII\ is apparently single. Therefore, the \logd\ values for 
       \LSXI\ have to be increased by $\approx\log\sqrt{2} = 0.151$. This implies that the derived distances for \LSXI\ and \LSXII\ are incompatible. Another 
       way to look at the discrepancy is that if we use the extinction-corrected apparent Johnson $V$ magnitude ($V_{J,0}$) for \LSXI\ and apply a 
       $\approx 2.5\log 2 = 0.753$ correction, we end up with two stars with $V_{J,0}$ values between 6.1 and 6.2, not brighter but actually slightly fainter than 
       \LSXII. In other words, if the three stars are at the same distance, we would require the two objects with spectroscopic class I to be fainter than the 
       object with spectroscopic class V. We analyze the distance issue later on. 
\end{itemize}

\subsection{Visual multiplicity}

\begin{table*}
\centerline{
\begin{tabular}{lrccc}
\hline
Pair & \multicolumn{1}{c}{Separation} & Orientation & $\Delta i$   & $\Delta z$      \\
     & \multicolumn{1}{c}{(\arcsec)}  & (degrees)   & (mag)        & (mag)           \\
\hline
\LSXI\ AB  & 22.601$\pm$0.054 & 139.94$\pm$ 0.18 & 5.263$\pm$0.011 & 5.209$\pm$0.012 \\
\LSXI\ AC  & 14.274$\pm$0.061 & 114.23$\pm$ 0.15 & 6.442$\pm$0.017 & 6.260$\pm$0.014 \\
\LSXI\ AD  & 23.803$\pm$0.023 & 131.77$\pm$ 0.07 & 6.888$\pm$0.037 & 6.445$\pm$0.040 \\
\LSXI\ AE  & 17.446$\pm$0.039 & 123.09$\pm$ 0.34 & 8.123$\pm$0.062 & 7.571$\pm$0.036 \\
\LSXI\ AF  & 12.300$\pm$0.056 & 105.54$\pm$ 0.22 & 8.381$\pm$0.266 & 7.802$\pm$0.226 \\
\LSXII\ AB &  6.114$\pm$0.050 & 318.83$\pm$ 0.20 & ---             & 6.968$\pm$0.034 \\
\LSXII\ AC & 11.135$\pm$0.050 & 318.60$\pm$ 0.20 & ---             & 7.781$\pm$0.053 \\
\hline
\end{tabular}
}
\caption{AstraLux results for \LSXI\ and \LSXII.}
\label{astralux}
\end{table*}

\begin{table*}
\centerline{
 \begin{tabular}{lcr@{}c@{}lr@{}c@{}lr@{}c@{}lc}
\hline
Star      & 2MASS ID          & \multicolumn{3}{c}{$J$}   & \multicolumn{3}{c}{$H$}   & \multicolumn{3}{c}{$K_{\rm s}$} & Flag \\
          &                   & \multicolumn{3}{c}{(mag)} & \multicolumn{3}{c}{(mag)} & \multicolumn{3}{c}{(mag)}       &      \\
\hline
\LSXI\ A  & J20351264+4651121 &  7.653&$\pm$&0.023        &  7.194&$\pm$&0.018        &  6.971&$\pm$&0.023              & AAT  \\
\LSXI\ B  & J20351402+4650549 & 12.326&     &             & 12.050&$\pm$&0.054        & 11.821&$\pm$&0.053              & UAA  \\
\LSXI\ C  & J20351389+4651065 & 13.402&$\pm$&0.049        & 12.847&$\pm$&0.066        & 12.585&$\pm$&0.070              & AAE  \\
\LSXI\ D  & J20351436+4650562 & 12.614&     &             & 12.638&$\pm$&0.033        & 12.235&$\pm$&0.030              & UAA  \\
\LSXI\ E  & J20351405+4651025 & 14.295&$\pm$&0.062        & 13.373&$\pm$&0.049        & 12.990&$\pm$&0.050              & AAA  \\
\LSXI\ F  & undetected        &       &     &             &       &     &             &       &     &                   &      \\
\LSXII\ A & J20351857+4650028 &  7.934&$\pm$&0.021        &  7.618&$\pm$&0.017        &  7.470&$\pm$&0.023              & AAA  \\
\LSXII\ B & J20351823+4650072 & 10.333&     &             & 10.158&     &             & 12.551&$\pm$&0.205              & UUC  \\
\LSXII\ C & J20351786+4650112 & 14.450&$\pm$&0.124        & 13.466&$\pm$&0.134        & 13.242&$\pm$&0.050              & BBA  \\
\hline
\end{tabular}
}
\caption{2MASS photometry for the targets detected in the AstraLux images. ABC flags are used for normal detections (with increasingly 
larger uncertainties), U flags for upper limits on magnitudes, and T for measurements done in this work.}
\label{twomass}   
\end{table*}

$\,\!$\indent Table~\ref{astralux} gives the separation, orientation and magnitude difference between the brightest star in the AstraLux Norte images (either
\LSXI\ or \LSXII) and the rest of the detected point sources. Note that the AstraLux images are not absolutely calibrated, so only differential photometry is
provided; hence, all the data refer to stellar pairs. Nevertheless, the photometry in the $i$ and $z$ bands for the two bright stars can be derived from the 
best SED in the previous subsection: \LSXI\ A has zero-point-corrected AB $i$ and $z$ magnitudes of 9.549 and 9.017, respectively, and \LSXII\ A has 
zero-point-corrected AB $i$ and $z$ magnitudes of 9.371 and 9.043, respectively. Table~\ref{twomass} gives the 2MASS $JHK_{\mathrm{S}}$ magnitudes for the stars
in the AstraLux images. All are detected except for \LSXI\ F, the dimmest star in the AstraLux image. Note that the 2MASS magnitudes are not complete, as
usual for moderately crowded fields such as this one. 

The most relevant result is that \LSXI\ has no visual companions within 11\arcsec\ and that \LSXII\ has just one dim companion 6\arcsec\ away. From the
photometric point of view, this means that the analysis in the previous subsection does not appear to include additional stars, so one can refer to the
photometry of \LSXI\ and \LSXI\ A as indistinguishable\footnote{\LSXI\ A includes two spectroscopic components but their separation, as previously derived, is of
the order of 1 mas, two orders of magnitude below what can be resolved with AstraLux Norte. See Fig.~2 in \citet{Maiz10a} for the values of separations and magnitude
differences that AstraLux Norte can resolve.} (as we have done in the previous paragraph) and the same can be said 
about \LSXII\ and \LSXII\ A. From the physical point of view, this means that (barring any undetected components) \LSXI\ is a double system but likely not a
higher-order one, since F or C (the closest companions) are likely to be unbound, especially considering that their environment is the center of a cluster. 
On the other hand, \LSXII\ B is $\sim 10\,000$ AU in the plane of the sky away from A and could possibly be bound \citep{Maiz10a}. There is also another 
component visible in the 2MASS images (but not listed in Tables~\ref{astralux}~or~\ref{twomass} because it fell just outside the 
$25\arcsec\times 25\arcsec$ AstraLux Norte field of view) 7\arcsec\ to the E of \LSXII\ A, thus increasing the probability of the existence of a bound 
companion.

\subsection{\BXC\ photometry}

\begin{figure}
\centering
\resizebox{\columnwidth}{!}{\includegraphics[clip]{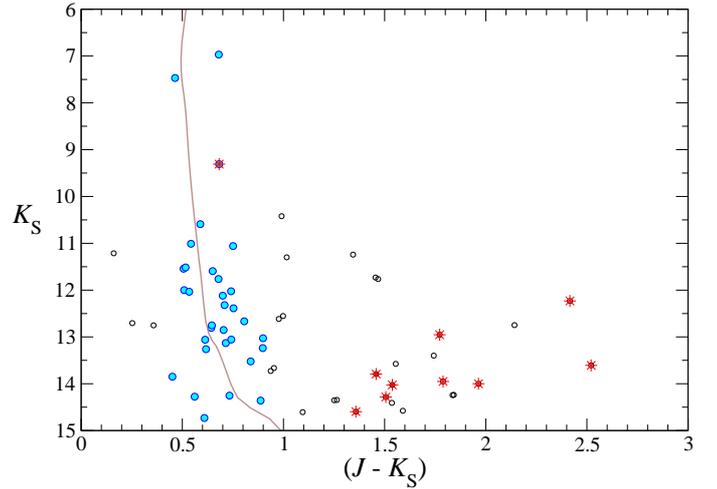}}
\caption{2MASS CMD for stars within $3\arcmin$ of the nominal center of \BXC\ passing the cuts described in the text. The blue circles are all the stars 
with $0.45<(J-K_{\mathrm{S}})<0.9$. Red-star symbols are used to mark stars with H$\alpha$ emission, as selected from Fig.~\ref{iphas}. 
The thick line is a 2~Ma isochrone with high rotation from \citet{Ekstetal12} displaced to $DM=12.0$ (\logd\ = 3.4) with $E(J-K_{\mathrm{S}})=0.75$.
The brightest objects are \LSXI\ (to the right of the isochrone) and \LSXII\ (to the left).} 
\label{2masscmd}
\end{figure}

$\,\!$\indent The area of \BXC\ is immersed in bright nebulosity (Sh~2-115, \citealt{Shar59}), indicating the existence of sources with large ionizing fluxes, 
but the cluster has surprisingly received very little attention. We have used photometric data from 2MASS 
\citep{Skruetal06} and IPHAS \citep{Bareetal14} to study its properties. We have selected 2MASS sources within $3\arcmin$ of the nominal center of the cluster,
as given in SIMBAD. We have rejected stars with bad quality flags, and cross-matched our selection with the IPHAS catalogue. As the IPHAS catalogue contains 
many spurious sources in areas of bright nebulosity, we have only accepted sources with a 2MASS counterpart within a $0\farcs6$ radius. The $Q_{{\rm IR}}$ 
parameter, defined as $Q_{{\rm IR}}=(J-H)-1.8\times(H-K_{\mathrm{S}})$, is very effective at separating early and late-type stars 
\citep[e.g.,][]{ComePasq05,NeguSchu07}, with early-type stars showing values $\approx0.0$. We select objects with $Q_{{\rm IR}}<0.08$ as candidate early-type 
stars \citep[see][]{NeguSchu07}, though emission-line stars also display negative values due to their $K_{\mathrm{S}}$ excesses. The candidates are clearly 
concentrated towards the cluster center, confirming that they mainly represent the cluster population. 
The $K_{\mathrm{S}}/(J-K_{\mathrm{S}})$ diagram for the resulting selection is displayed in Fig.~\ref{2masscmd}.

\begin{figure}
\centering
\resizebox{\columnwidth}{!}{\includegraphics[clip]{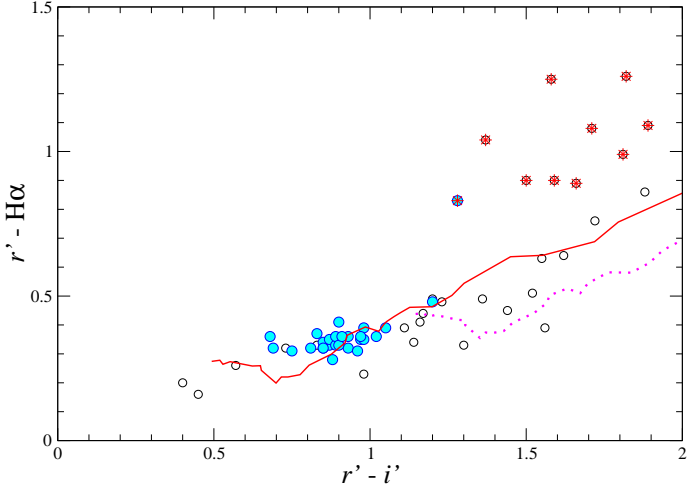}}
\caption{IPHAS $(r'-i')$/$(r'-\mathrm{H}\alpha)$ diagram. The solid (red) line is the locus of the main sequence for $\ebv =1$ from \citet{Drewetal05}, 
while the (magenta) dotted line is the locus of the main sequence for $\ebv=2$, roughly defining the envelope for the hot stars that are cluster members. 
Blue circles are used to indicate the stars selected from Fig.~\ref{2masscmd} and marked in the same way there. Since all 
the stars have been selected to have a pseudocolor $Q_{\mathrm{IR}}$ typical of early-type stars or emission-line stars and they all have relatively red
$(J-K_{\mathrm{S}})$ colors, we can safely assume that stars clearly above the main-sequence locus are emission-line stars. Their positions in the
2MASS CMD, (marked with red-star symbols both here and in Fig.~\ref{2masscmd}), are in all cases compatible with this interpretation. Note that both \LSXI\ and
\LSXII\ are saturated in IPHAS, as all bright O stars in the northern sky are.}
\label{iphas}    
\end{figure}

Possible cluster members show a broad distribution in $(J-K_{\mathrm{S}})$. Three objects with $(J-K_{\mathrm{S}})<0.4$ are located away from the cluster and 
may be foreground stars. Interestingly, \LSXII\ has the second lowest $(J-K_{\mathrm{S}})$ of all possible members, with $0.46\pm0.03$, while the location of 
\LSXI\ in Fig.~\ref{2masscmd} shows that it is more reddened (as we already knew from the CHORIZOS analysis). We assume that the bulk 
of cluster members is given by the vertical strip extending between $(J-K_{\mathrm{S}})=0.45$ and $(J-K_{\mathrm{S}})=0.9$. This is confirmed by their 
concentration in the IPHAS $(r'-i')$/$(r'-\mathrm{H}\alpha)$ diagram (Fig.~\ref{iphas}), where all except three are distributed in a narrow strip with 
$0.7\la (r'-i')\la 1.05$ that follows the reddening vector, and the vast majority have $0.8\la (r'-i')\la 1.0$. The main concentration of cluster members, lying 
between \LSXI\ and \LSXII, shows only a small spread in color, while the stars lying immediately adjacent to \LSXI\, both to the North and West display higher 
values. The average $(J-K_{\mathrm{S}})$ for all stars with values between 0.45 and 0.9 is 0.67, with a standard deviation $\sigma=0.12$, showing that the 
objects are evenly distributed between these values and therefore defining a typical reddening for the cluster is meaningless. 

A significant number of sources occupy positions in the $(r'-i')$/$(r'-\mathrm{H}\alpha)$ diagram compatible with emission-line stars, displaying 
$(r'-\mathrm{H}\alpha)>0.8$ and $(r'-i')>1.2$ \citep{Corretal08}. Almost all these objects are quite faint and display high values of $(J-K_{\mathrm{S}})>1.4$. 
More than half have $Q_{{\rm IR}}<-0.1$, typical of emission line stars. These objects can represent a population of pre-main sequence stars associated to the 
cluster. Interestingly, none of them is located in the central region of the cluster.

As there are no obviously evolved stars, it is not possible to assign an age to \BXC. Note that the luminosity difference between dwarfs and supergiants is 
relatively small at the earliest spectral types. Assuming that \LSXII\ is slightly evolved, we 
can perhaps give an age $\sim2$~Ma, but luminous O-type stars in Cygnus~OB2 with spectral types O3-O5 are significantly hotter than the main-sequence turn-off 
\citep{Neguetal08a}. By analogy, an age up to 3~Ma is possible. For illustration, Fig.~\ref{2masscmd} shows a high-rotation isochrone from \citet{Ekstetal12} 
corresponding to an age of 2~Ma, reddened by a representative  $E(J-K_{\mathrm{S}})=0.75$ and displaced to $DM=12.0$~mag (\logd ~=~3.4). The fit to the position of 
objects in the central concentration is rather good, considering the variable reddening. Three objects lie well to the left of 
the isochrone and therefore are either non-members or outliers with low reddening. All of them lie to the East of \LSXII, except one that is located at the 
northern edge of the region analyzed. However, removing these objects does not change the average color or its standard deviation. On the other hand, a few 
objects lying to the right of the isochrone, with $(J-K_{\mathrm{S}})\approx1.0$ are likely cluster members with higher than average reddening.

In spite of the presence of three early O-type stars (two in \LSXI\ and at least one in \LSXII), \BXC\ seems to contain very few OB stars. Only one photometric 
member is sufficiently bright in $K_{\mathrm{S}}$ to be a late-O star, 2MASS~J20350798+4649321 (Fig.~\ref{2massimage}), and this object has a position in the 
IPHAS diagrams consistent with being an emission-line star (it is the object marked with both a blue circle and a red-star symbol in 
Figs.~\ref{2masscmd}~and~\ref{iphas}). The bulk of the population starts almost 3~magnitudes below \LSXII, at $K_{\mathrm{S}}\approx10.5$, an intrinsic magnitude 
roughly corresponding to a B1~V spectral type.

\section{Discussion}

$\,\!$\indent O stars earlier than type O4 are very scarce in the Galaxy. Prior to this work, there were only two examples known in the northern hemisphere,
Cyg~OB2-7 and Cyg~OB2-22~A \citep{Walbetal02b,Sotaetal11a}. Given that we know so few very massive stars, it is crucial to keep searching for them to increase our 
statistics if we want to establish what is the stellar upper mass limit and its dependence on metallicity and environmental conditions. 

We have shown that \LSXI\ as a very massive eccentric binary composed of two near-twin stars. A few years ago this may have been seen as a fluke but several 
similar systems have been discovered recently. 

\begin{itemize}
 \item HD~93\,129~AaAb \citep{Nelaetal04,Nelaetal10,Maizetal05b,Maizetal08b,Sotaetal14} may be even more massive, given that the primary is of spectral type O2~If*, 
       but the secondary is one magnitude fainter and the orbit is yet undetermined and appears to be decades to centuries long. 
 \item Cyg OB2-9 (\citealt{Nazeetal12c}, O5-5.5~I~+~O3-4~III) has a very similar mass ratio and a somewhat larger eccentricity, its $(M_1+M_2)\sin^3 i$ is only 
       slightly lower, its period is an order of magnitude larger, and the spectral types are slightly later. 
 \item R139 (\citealt{Tayletal11a}, O6.5~Iafc~+~O6~Iaf) is also similar to \LSXI\ in terms of period, eccentricity, and mass ratio but the two stars are mid-O 
       supergiants and their lower mass limits are significantly higher, between 60 and 80 M$_\odot$. Those masses indicate that not all objects above 60 M$_\odot$ 
       are WNh stars (of course, as long as we do not know the inclinations in this and other cases, we will not know what the true masses are). 
 \item Two additional examples of very massive stars in elliptical orbits whose large eccentricity made the discovery of their binarity difficult are WR~22 $\equiv$
       HD~92\,740 (\citealt{MoffSegg78,Contetal79,Schwetal99}, WN7~+~O8-9.5~III:) and HD~93\,162 (\citealt{Gameetal08b,Sotaetal14}, O2.5~If*/WN6~+~OB). 
 \item There are also very massive twin systems in shorter, near-circular orbits such as NGC~3603-A1 (\citealt{Moffetal04,Schnetal08a}, WN6ha~+~WN6ha), WR~20a 
       (\citealt{Rauwetal04,Bonaetal04,CrowWalb11}, O3~If*/WN6~+~O3~If*/WN6), and Pismis 24-1 (\citealt{Maizetal07}, O3.5~If*~+~O4~III(f)~+~\ldots). The latter also 
       includes a third very massive component in a long orbit.
 \item All of those systems are located in regions with large numbers of O stars (Carina~Nebula; Cygnus~OB2; 30~Doradus; NGC~3603; Westerlund~2; and, to a 
       lesser extent, Pismis~24). \LSXI\ is the oddity in that respect, being located in a significantly less massive cluster or association. In that respect, 
       a more similar case may be HD~150\,136, which is apparently less evolved but is a triple system [O3-3.5~V((f*))~+~O5.5-6~V((f))~+~O6.5-7~V((f))] with a most 
       massive star of 53$\pm$10~M$_\odot$ in a relatively small cluster, NGC~6193, with another nearby star, HD~150\,135 [O6.5~V((f))z], yielding a similar makeup 
       to that of \BXC\ \citep{NiemGame05,Mahyetal12,Sanaetal13a,SanBetal13,Sotaetal14}.
\end{itemize}

All of the cases mentioned above refer to distant very massive stars (Cyg OB2-9 and HD~150\,136 are the closest ones but they are beyond 1 kpc) but
what should be even more surprising is that some of the brightest and closest O stars in the sky have been recently discovered to have eccentric companions. 
That is what has happened with $\theta^1$~Ori~C (\citealt{Krauetal07,Sotaetal11a}, O7~Vp~+~\ldots) and $\sigma$~Ori~A (\citealt{SimDetal11c,SimDetal15a}, 
O9.5~V~+~B0/1~Vn, note that $sigma$~Ori~B is B0.5~V). Why have not these systems been 
discovered before? There are two reasons: eccentric systems require extensive spectroscopic monitoring since their velocity differences may be too small to be 
detected during a large fraction of their orbits (as it happened with \LSXI) and in some cases interferometry is the only way to detect their multiplicity because
of their large semi-major axes. It has not been until the last 
decade that large-scale spectroscopic monitoring of many O stars has started and that interferometric technology has allowed similar surveys using those techniques. 
However, many systems still remain outside the reach of such surveys so discoveries should continue in the following years.

Our data cannot provide conclusive results on the masses of \LSXI\ and \LSXII. Without eclipses, we cannot accurately measure the inclination of the \LSXI\ orbit and our 
keplerian masses of 38.80$\pm$0.83~M$_\odot$ and 35.60$\pm$0.77~M$_\odot$ have the $\sin^3 i$ factor included in them, making them just lower limits. Note that
the star is not present in the public release of the Northern Sky Variability Survey \citep{Woznetal04} and that the time coverage in the SuperWASP 
\citep{Polletal06} public archive is very limited, so a thorough search for eclipses is not possible at this time. In any case, eclipses
are unlikely, since the large separations in the orbit would require an inclination very close to 90 degrees in order for them to take place. 
The inclination could be constrained in the future with broadband polarimetry or through the phase-dependent behaviour of excess emission from colliding winds

For evolutionary masses, we have a different problem: our results for the spectral types and distances are inconsistent. There are three possible explanations:

\begin{enumerate}
 \item A straightforward interpretation of the CHORIZOS results place \LSXI\ at a \logd\ between 3.45 and 3.67 (in parsecs, after correcting for the existence of 
       two stars) and \LSXII\ at a \logd\ between 3.10 and 3.28. Therefore, one possibility would be that both objects are physically unrelated and just the
       result of a chance alignment. In this case, the luminosity classification criteria for early-type O stars would retain their physical meaning and there
       would be no need to invoke the existence of undetected multiple systems. Nevertheless, we judge this possibility to be highly unlikely, given the small 
       number of early-type O stars that exist and the existence of the underlying cluster \BXC. In any case, this hypothesis will be tested soon once the Gaia 
       parallaxes become available. 
 \item An alternative would place both \LSXI\ and \LSXII\ at a \logd$\sim$3.4, consistent with the 2MASS photometry of \BXC, but would require that the current
       luminosity classification criteria for O stars based on the depth of \HeII{4685.71} does not really reflect a function or luminosity but instead is just a 
       measurement of wind strength\footnote{Also note that the continuum of \LSXII\ to the right of \HeII{4685.71} is slightly rised, possibly signalling an anomalous 
       profile indicative of luminosity class III rather than V, which would have a similar effect to the one described here.}. We consider this option unlikely, as 
       there are both theoretical reasons why (for the same \teff\ and metallicity) wind strength should strongly correlate with luminosity and observational data that 
       corroborate that association (e.g. \citealt{Walbetal14}). This hypothesis could be tested by obtaining a good-S/N high-resolution spectrum of \LSXII\ and 
       modelling it with e.g. FASTWIND or CMFGEN. Those codes derive gravity from a fit to the Balmer line profiles which is independent of \HeII{4685.71}.
 \item A third option is that \LSXII\ is another near-twin binary system composed of two very-early-type O dwarfs. In this scenario, \logd\ would be in the
       3.40-3.45 range, marginally consistent with the spectroscopic parallaxes for \LSXI\ and \BXC. This solution may be ad hoc but we believe it to be the most
       likely one. Indeed, there is a precedent with an object that is a near-spectroscopic twin of \LSXII, HD~93\,250 (O4~III(fc), \citealt{Sotaetal14}). Its 
       spectroscopic parallax was incompatible with the well known value of the Carina Nebula until it was discovered to be a binary through interferometry 
       \citep{Sanaetal11b}. Note that there is a large range of separations for which we would not detect significant velocity variations if the system were a 
       spectroscopic binary, that the inclination could be small, that our spectroscopic campaign has not been as thorough for \LSXII\ as it has been for \LSXI, 
       and that X-ray excesses due to wind-wind collisions are expected to be weaker for dwarfs than for supergiants. 
\end{enumerate}

Related to the issue of the masses is the age of \BXC. Until one of the above explanations is confirmed, we cannot provide a final answer. We should note,
however, that if the third option is the correct one, it is possible to derive a consistent age of 1.5-2.0~Ma for \LSXI, \LSXII, and \BXC\ using the low-rotation
Geneva evolutionary tracks of \citet{LejeScha01}. Under those assumptions, the masses of the two stars in \LSXI\ would be in the 60-70~M$_\odot$ range and those in 
\LSXII\ would be 35-45~M$_\odot$.

The results in this paper reveal that \BXC\ can be an important cluster to resolve the issue of how the stellar Initial Mass Function (IMF) is sampled. There are two
alternative theories, one that states that the IMF is built in a sorted way, with many low-mass stars being formed before any high-mass star can appear, and another one 
that states that the IMF is built in a stochastic way, with some rare cases where it is possible to form high-mass stars with a relatively small number of low-mass
ones (see \citealt{Bresetal12} and references therein). If the third option on the \LSXI\ and \LSXII\ masses above is correct, we would have a cluster with four stars 
above 35~M$_\odot$, one of them with $\sim$70~M$_\odot$, and none or just one (2MASS~J20350798+4649321) 
between 20~and~35~M$_\odot$ (see Fig.~\ref{2masscmd} and the associated discussion in the text). Using the total stellar mass above 20~M$_\odot$ and assuming a Kroupa 
IMF \citep{Krou01} between 0.1~M$_\odot$ and 100~M$_\odot$, we derive that the mass of \BXC\ is $\approx$2000~M$_\odot$, 
which is consistent with the appearance of the cluster in relationship with other similar objects. However, such a large maximum stellar mass in such a small cluster
could be incompatible with the sorted sampling scenario \citep{WeidKrou06} but can be accomodated within the stochastic scenario: a 2000~M$_\odot$ with the conditions above
should have, on average, 2.3 stars in the 20-35~M$_\odot$ range and 1.6 stars in the 35-100~M$_\odot$ range. We observe 1 and 4, respectively i.e. off from the expected values
but within the range of possibilities expected for a single case within the number of similar clusters in the solar neighborhood. We need better
data (constraining the nature of \LSXII\ with interferometry, studying a deep CMD of \BXC\ to accurately measure the cluster mass, and resolving the distance issues with 
Gaia) to provide a final answer but \BXC\ appears to be a good candidate for a small cluster with a star too massive for the sorted sampling scenario.

\section{Conclusions}

\begin{itemize}
 \item \BXC\ is a young stellar cluster dominated by two early O-type systems. \LSXI\ is an SB2 composed of two very similar O3.5~If* stars. \LSXII\ is spectroscopically
       single and has a spectral type O4.5~V((f)).
 \item \LSXI\ has an eccentric orbit ($e$ between 0.56 and 0.57) with a 97.2-day period and minimum masses of 38.80$\pm$0.83~M$_\odot$ and 35.60$\pm$0.77~M$_\odot$. 
       Since we do not know the inclination we cannot calculate accurate keplerian masses.
 \item \LSXI\ has a significantly higher extinction than \LSXII. The optical+NIR extinction law is close to the average one in the Galaxy.
 \item There are no apparent bright visual companions to either system.
 \item The evolutionary masses of \LSXI\ and \LSXII\ are incompatible with the two systems being located at the same distance and having the same age. We consider different
       solutions to the problem and consider that the most likely one is the existence of an undetected companion for \LSXII, for which there is plenty of room in terms of period, 
       inclination, and eccentricity not yet explored.
 \item \BXC\ is a cluster with considerable differential extinction and its stellar mass is possibly too low to harbor both \LSXI\ and \LSXII\ under the sorted sampling scenario
       for the IMF. 
\end{itemize}

\begin{acknowledgements}
We thank the referee, Tony Moffat, for his useful comments that helped improve this paper.
J.M.A. and A.S. acknowledge support from [a] the Spanish Government Ministerio de Econom{\'\i}a y Competitividad (MINECO) through grants 
AYA2010-15\,081, AYA2010-17\,631, and AYA2013-40\,611-P and [b] the Consejer{\'\i}a de Educaci{\'o}n of the Junta de Andaluc{\'\i}a through 
grant P08-TIC-4075. J.M.A. was also supported by the George P. and Cynthia Woods Mitchell Institute for Fundamental Physics and Astronomy and 
he is grateful to the Department of Physics and Astronomy at Texas A\&M University for their hospitality during some of the time this work was 
carried out. I.N., A.M., J.A., and J.L. acknowledge support from [a] the Spanish Government Ministerio de Econo\'{\i}a y Competitividad (MINECO) 
through grant AYA2012-39\,364-C02-01/02, [b] the European Union, and [c] the Generalitat Valenciana through grant ACOMP/2014/129.
R.H.B. acknowledges support from FONDECYT Project 1\,140\,076.
S.S.-D. acknowledges funding by [a] the Spanish Government Ministerio de Econom{\'\i}a y Competitividad (MINECO) through grants 
AYA2010-21\,697-C05-04, AYA2012-39\,364-C02-01, and Severo Ochoa SEV-2011-0187 and [b] the Canary Islands Government under grant PID2\,010\,119.
J.S.-B. acknowledges support by the JAE-PreDoc program of the Spanish Consejo Superior de Investigaciones Cient\'{\i}ficas (CSIC).
STScI is operated by the Association of Universities for Research in Astronomy, Inc., under NASA contract NAS5-26555.
\end{acknowledgements}

\bibliographystyle{aa}
\bibliography{general}

\end{document}